\documentclass[aps,prl,superscriptaddress,twocolumn,showpacs,floatfix]{revtex4}

\usepackage{amsmath}
\usepackage{amssymb}
\usepackage{graphicx}
\usepackage{mathptmx} 
\usepackage{color}  
\usepackage{bm}

\begin{document}

\title{Fermi Surface Reconstruction and Quantum Oscillations in Underdoped YBa$_2$Cu$_3$O$_{7-x}$ Modeled in a
Single Bilayer with Mirror Symmetry Broken by Charge Density Waves}

\date{\today}

\author{A. Briffa}
\affiliation{School of Physics and Astronomy, University of Birmingham, Birmingham B15 2TT, United Kingdom.}

\author{E. Blackburn}
\affiliation{School of Physics and Astronomy, University of Birmingham, Birmingham B15 2TT, United Kingdom.}

\author{S. M. Hayden}
\affiliation{H. H. Wills Physics Laboratory, University of Bristol, Bristol, BS8 1TL, United Kingdom.}

\author{E. A. Yelland}
\affiliation{SUPA, School of Physics and Astronomy, University of St Andrews, St Andrews, KY16 9SS, United Kingdom.}
\affiliation{SUPA, School of Physics and Astronomy and CSEC, University of Edinburgh, Edinburgh, EH9 3FD, United Kingdom.}

\author{M. W. Long}
\affiliation{School of Physics and Astronomy, University of Birmingham, Birmingham B15 2TT, United Kingdom.}

\author{E. M. Forgan}
\affiliation{School of Physics and Astronomy, University of Birmingham, Birmingham B15 2TT, United Kingdom.}

\begin{abstract}
Hole-doped high-temperature cuprate superconductors below optimum doping have small electron-like Fermi surfaces occupying a small fraction of the Brillouin zone. There is strong evidence that this is linked to charge density wave (CDW) order, which reconstructs the large hole-like Fermi surfaces predicted by band structure calculations . Recent experiments have revealed the structure of the two CDW components in the benchmark bilayer material YBa$_2$Cu$_3$O$_{7-x}$ in high field where quantum oscillation (QO) measurements are performed. We have combined these results with a tight-binding description of the bands in an isolated bilayer to give a minimal model revealing the essential physics of the situation. Here we show that this approach, combined with the effects of spin-orbit interactions and the pseudogap, gives a good qualitative description of the multiple frequencies seen in the QO observations in this material. Magnetic breakdown through weak CDW splitting of the bands will lead to a field-dependence of the QO spectrum and to the observed fourfold symmetry of the results in tilted fields. 
\end{abstract}


\maketitle

\section{Introduction}
In recent years, a clear picture of the variation of electronic structure of cuprate high-Tc materials with doping has emerged. Overdoped materials have large hole-like cylindrical Fermi surfaces with cross sections reflecting the single hole on a Cu$^{2+}$ plus the additional carriers due to doping \cite{QO-od}, whereas on the under-doped side of the superconducting dome, QO measurements \cite{QO2007} indicated small electron-like Fermi surface (FS) areas that occupy only $\sim$ 2 \% of the Brillouin zone (BZ). This strongly suggested a FS reconstruction arising from broken translational symmetry \cite{LeBoeufHall2007}. Macroscopic measurements indicate a change to electron-like transport below $\sim$ 150 K in the under-doped region \cite{Chang2010}, which is consistent with the FS reconstruction scenario, and NMR measurements on under-doped YBa$_2$Cu$_3$O$_{7-x}$ (YBCO) \cite{Wu2011} give clear indications of charge density waves. 
The CDW state has been seen by X-ray diffraction in YBCO \cite{Ghiringhelli, Achkar, Blanco-Canosa2013, Blanco-Canosa2014, Chang2012, Blackburn, Huecker}. It has been shown to be a ubiquitous High-Tc phenomenon \cite{Comin2014,SilvaNeto2014,Hashimoto2014,Thampy2014,Tabis2014,Christensen2014, Croft2014,Comin_arXiv,Fujita, SilvaNeto2015}, and there is strong evidence that the CDW order at high fields is intimately connected with the QO frequencies \cite{Tabis2014}. 

 In zero magnetic field, the CDW displacements break the mirror symmetry of the CuO$_2$ bilayers \cite{Forgan2015}.  In high magnetic fields, the CDW modulation along the crystal $b$-direction of an YBCO sample develops long-range order \cite{Gerber2015}.  The relationship of the high-field structure to the zero-field structure \cite{Forgan2015} has also been established \cite{Chang2015} and it also breaks the mirror symmetry of the CuO$_2$ bilayers. Hence we now have the essential ingredients to give an explanation of the QO results. Unlike previous attempts, in our model we use the correct CDW symmetry, and choose our chemical potential to be consistent with the nodal FS recently seen by ARPES. We assume weak $c$-axis tunnelling between bilayers in order to obtain the minimal model for the QO results. This approach reveals how much can be explained by a single-particle Fermi liquid model, and raises some fascinating questions.
\section{Basic mechanism for the FS reconstruction}
In the present paper, we consider the Fermi surface reconstruction of a single bilayer in which both \textbf{a} and \textbf{b} modulations are present in the same region of the sample.  This coexistence is consistent with ultrasonic measurements on an underdoped YBCO sample \cite{LeBoeuf2013} and the close-to-fourfold symmetry of the QO results \cite{SandH2014}. Recent zero-field X-ray measurements \cite{Forgan2015} have shown that both the CDWs have a previously unsuspected symmetry, which gives opposite perturbations in the two halves of the superconducting CuO$_2$ bilayers. X-ray measurements also show \cite{Chang2015} that this symmetry is maintained at high magnetic fields $B \approx 17$~T,  i.e. above the phase transition observed by ultrasound and NMR for fields B $\sim$15-17 T.  However,  x-ray measurements \cite{ Chang2012,Chang2015} indicate a very short $c$-axis correlation length for the CDW modulated along \textbf{a}, yet clear QO signals are observed in what is a somewhat disordered state.  Thus we believe that it is reasonable to consider a single bilayer occasioned by weak $c$-axis tunnelling between separate bilayers, which will be little affected by incomplete ordering along the $c$-axis. 

There is a hierarchy of electronic energy scales in this system, and we list the ones of relevance for this problem. The largest is the Coulomb energy, which ensures a Mott insulating state at zero doping. The next largest is the basal plane hopping, which gives the shape and the effective mass of the unreconstructed Fermi surface. There are then several effects which are of similar order of magnitude, and which we list in order of expected decreasing size: $c$-axis tunnelling between the two halves of a single CuO$_2$ bilayer, pseudogap energy (which we shall include only qualitatively) and the CDW perturbation. The energy of the carriers due to the applied magnetic field must also be included: in a typical field of 50~T used in QO measurements, the Zeeman energy of the spins is $\sim\pm$3 meV, and the splitting of the `orbital' Landau levels is similar. We expect that the spin-orbit interaction will be comparable in magnitude with this. Finally we have the $c$-axis tunnelling between separate bilayers, and we shall set this to zero in our model. 

In Fig.~1 we give a general outline of how the FS reconstruction arises from the CDW structures now established. The CDW perturbation can give strong hybridisation between states at the same energy which are connected by the CDW $\bf{q}$-vectors. The relevant states may be visualised by translating Fermi surfaces by these $\bf{q}$-vectors into a corner of the BZ - as illustrated in Fig.~1. Because the CDW gives opposite perturbations in the two halves of a bilayer, the strongest hybridisation takes place between the bonding ($B$) and antibonding ($A$) states of a bilayer, giving the avoided crossings at the points marked in black. However, weaker hybridisation can also take place at $A-A$ and $B-B$ crossings marked in red. This gives rise to two reconstructed FSs: the smaller one is of mainly $A$ character and the larger is mainly constructed from the $B$ bands. In high fields, the smaller gaps may be bridged by magnetic breakdown, giving several different FS areas, as observed in experiment. Parts of this general scenario have been invoked before \cite{Tabis2014, SandH2014, HandS2012, SHL2012, DL2015, HRS2015,MZRK2015}, but none of these models  used the correct form for the CDW perturbation, the tunnelling between the two halves of the bilayer, and the spin-orbit (SO) coupling at the FS crossing points.
\begin{figure}[t]
\begin{center}
\includegraphics[width=0.95\linewidth]{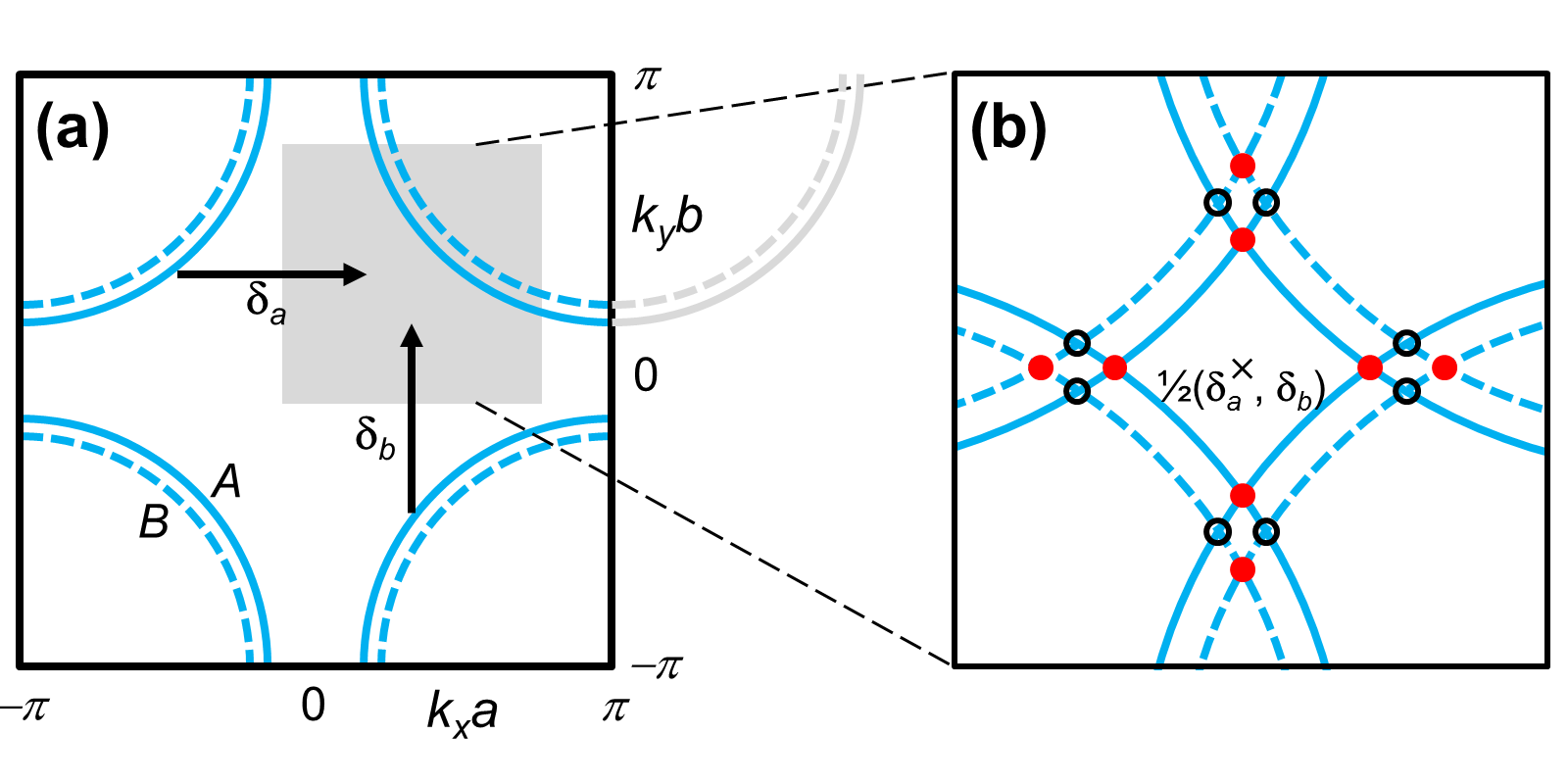}
\includegraphics[width=0.95\linewidth]{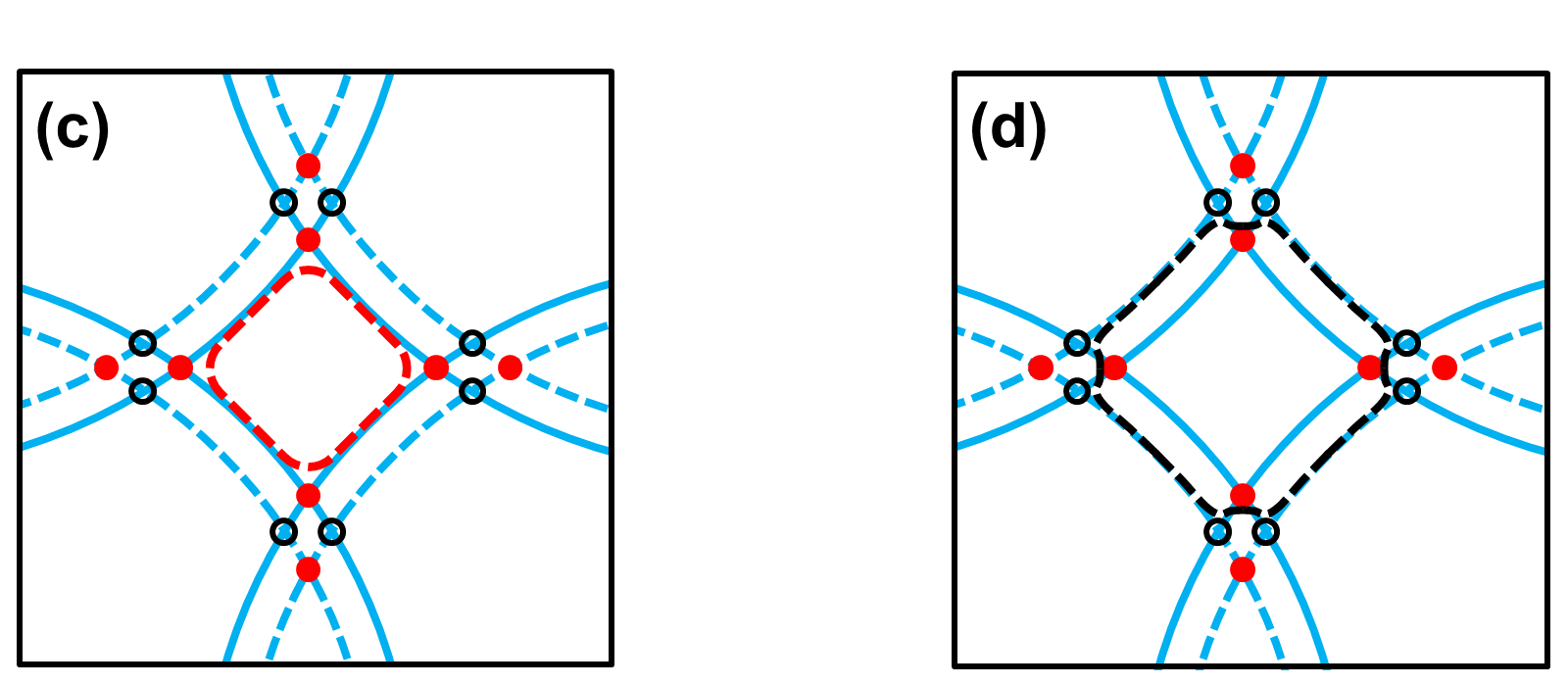}
\includegraphics[width=0.95\linewidth]{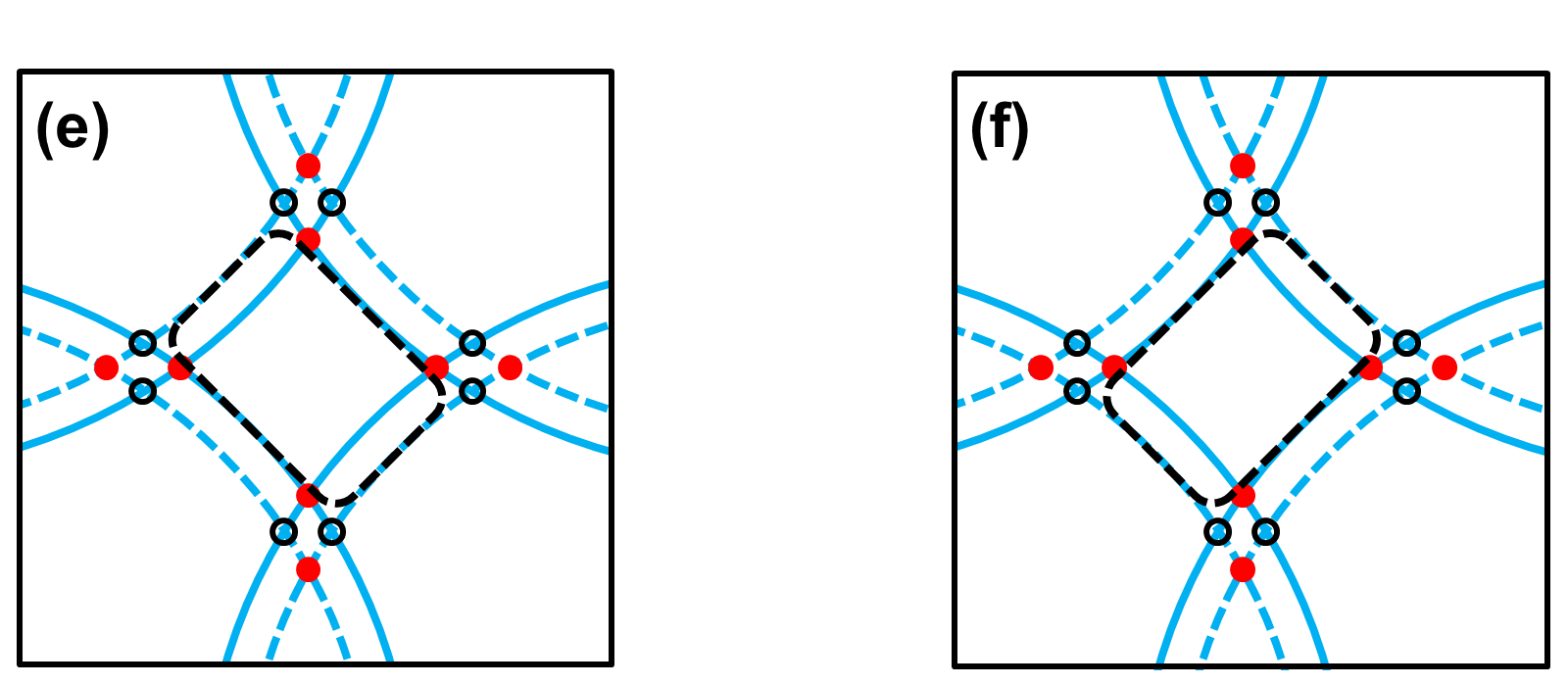}
\end{center}
\caption{Schematic illustration of CDW hybridisation of CuO$_2$ bilayer electron states. In (a) are represented the FS sheets arising from the antibonding ($A$) and bonding ($B$ - dashed line) bands of the CuO$_2$ bilayers in a $k_x-k_y$ cross-section of the BZ for YBCO. These bands form \emph{large} hole surfaces of approximately cylindrical shape around the corner of the BZ. The basal plane parts of the CDW wavevectors $\bm{\delta}_a$ and $\bm{\delta}_b$ are shown approximately to scale. The shaded region in (a), which is centred on the point $ \frac{1}{2}(\bm{\delta}_a, \bm{\delta}_b)$, is shown in the other 5 panels. The effects of the CDW components may be represented by translation of the bands in the other corners of the BZ by $\bm{\delta}_a$ and $\bm{\delta}_b$, giving crossings between states connected by one of the CDW wavevectors. The positions where the $A$ and $B$ states are degenerate and $A-B$ hybridisation may occur are marked with open black circles. At the red dots, there is $A-A$ and $B-B$ degeneracy, and weaker hybridisation occurs. The resulting small electron-like FS pockets are shown in red and black. If the hybridisation at the $A-A$ crossings is weak, , it can be crossed at high magnetic fields, and new FS areas, including those represented in (e) and (f) will result.}
\end{figure}
For optimally and underdoped cuprates, it is particularly clear that strong correlation effects, arising from the proximity to a Mott insulator state are important. These are associated with the `pseudogap' (PG) which corresponds to the removal of states at the Fermi level, leaving `Fermi arcs' instead of complete Fermi surfaces.  These have been observed in underdoped YBCO by photoemission (ARPES) \cite{Fournier2010}. The states removed by the PG are at the edges of Figure 1 and do not participate in the formation of the small electron-like pockets. It is also found \cite{Fournier2010} that the states in the CuO chains are suppressed. QO observations in YBCO as a function of underdoping confirm this effect, as they give no indications of a chain FS (nor effects of chain-ordering), and heat capacity measurements at high field \cite{Riggs2011} also show that the chains are not metallic. These results indicate that first-principles Density Functional Theory (DFT) calculations are not sufficiently accurate in the underdoped region, since they give a chain Fermi surface, and certainly do not give the PG. 

Instead, to obtain tractable results, we carry out our calculations using a tight-binding approximation (TBA) for the band structure of a single CuO$_2$ bilayer, which can be used to give the shapes of the Fermi arcs, which are coupled by the CDW. This approach lays bare the essential physics, because the symmetry of the band couplings caused by the CDW perturbations will be unaffected by the PG. In our two-dimensional model, we can ignore $k_z$ - the $c$-axis component of the carrier wavevector, but our results can be extended \cite{BriffaTBP} to take account of $c$-axis coherence revealed by transport measurements \cite{Vignolle2012}.

We first recount the properties of the bilayer states in the absence of a CDW. We then show how the hybridisation of bilayer states by a CDW varies with their bonding or antibonding character, the bilayer coupling, and the presence of spin-orbit splitting. We use the results of DFT calculations to indicate general features of bilayer coupling in the TBA approach, and reveal some misconceptions in the literature about this. We show results of our calculations for a set of parameters chosen to reveal the important features of the model that can give an account of features of the QO observations. We estimate the effects of magnetic breakdown across the CDW band-gaps and show that this will be important. Finally we discuss to what extent this scenario can account for the experimental observations.
\section{The nature of the bilayer wavefunctions and effects on CDW matrix elements} 
Until now it had been assumed that a CDW in YBCO would have even symmetry about the centre of the bilayer and would only couple $A$ to $A$ and $B$ to $B$ bands \cite{HandS2012,SandH2014}. However, structure determinations show that the CDW perturbation has odd symmetry, at both zero and high fields \cite{Forgan2015, Chang2015}. This perturbation may arise from CDW-associated changes in basal plane tunnelling, local doping or Coulomb potential, and all of these have opposite values in the two halves of a bilayer. In an isolated bilayer, the $A$ and $B$ states are usually assumed to have a definite parity so that the CDW perturbation couples only $A$ to $B$.  However, two effects negate this simple picture and allow weak hybridisation of $A$ with $A$ states and $B$ with $B$. Firstly, a Rashba-type spin-orbit interaction \cite{Rashba} gives an opposite energy shift to states of a given spin in the top and bottom layers. The importance of this interaction, and the fact that it can flip carrier spin has recently been emphasised \cite{HRS2015}.  Secondly, if coupling between adjacent bilayers cannot be ignored, the $A$ and $B$ states do not have a definite parity at general $k_z$. In the present Paper, we shall assume that this second effect can be neglected and consider this generalisation in a further publication \cite{BriffaTBP}.  


To set the scene, we sketch the TBA for the double Fermi surface sheets arising from the CuO$_2$ bilayer bands, ignoring the spin-orbit interaction for now. As described above, we also ignore the chain bands and for simplicity treat the YBCO cell as tetragonal with lattice parameters $a$ and $c$. In the absence of bilayer coupling, the energy $\epsilon({\bf{k}})$ of a state relative to the Fermi energy $\epsilon_F$ in a single CuO$_2$ layer with basal plane wavevector $\bf{k} = (k_x, k_y)$ is given by Ref. \onlinecite{Andersen}:
\begin{eqnarray}
\epsilon({\bf{k}}) = -2t (\cos k_x a + \cos k_y a) + 4 t' (\cos k_xa \cos k_ya) \nonumber\\
 - 2t'' (\cos 2k_xa + \cos 2k_ya) - \epsilon_F
\end{eqnarray}
Typical ratios for the tunnelling matrix elements are: $t' = 0.32 t$ and $t'' = 0.16 t$ \cite{Andersen}, and the value of $\epsilon_F$ is chosen to give an almost half-filled BZ. This $\epsilon({\bf{k}})$ is then modified by the intra-bilayer tunnelling $t_{\perp}$ and the coupling $t_c$ of two adjacent bilayers in the $c$-direction to give the two bilayer bands - bonding - of lower energy, and antibonding. We expect that $t_{\perp}$, $t'$, $t''$ and also the CDW perturbation $V$ will be of similar magnitude: in the region of 10 meV \cite{Chang2012}. The bilayer energies and states in the absence of the CDW are given by a Hamiltonian matrix using the basis states on each half of a bilayer \cite{DGA+SC}:
\begin{equation}
    H_b({\bf{k}}, k_z) = \begin{bmatrix}
        \epsilon({\bf{k}}) & -t_{\perp}-t_c\mathrm{e}^{-ik_zc} \\
        -t_{\perp}-t_c\mathrm{e}^{+ik_zc} & \epsilon({\bf{k}})
     \end{bmatrix}.
\end{equation}
The energy eigenvalues of this are:  
\begin{equation}
\epsilon_{A,B}({\bf{k}},k_z) = \epsilon({\bf{k}}) \pm \sqrt{t_{\perp}^2 +t_c^2 + 2 t_{\perp} t_c \cos k_z c},
\end{equation}
Assuming that $t_c$ is negligible, this reduces to $\epsilon_{A,B}({\bf{k}}) = \epsilon({\bf{k}}) \pm t_{\perp}$, where the $\pm$ ambiguity refers to the $A$ and $B$ bands respectively.  Note that the $c$-axis dispersion is removed if either of $t_{\perp}$ and $t_c$ is zero, but the $A-B$ splitting remains as long as at least one of them is nonzero.

We now introduce the spin-orbit (SO) interaction; there are several possible terms, but the important one for our purposes breaks the symmetry between the two halves of the bilayer for a state of given spin, and hence allows hybridisation by the CDW at crossings between bilayer states of like character. It arises because an individual CuO$_2$ layer does not have mirror symmetry in the $z$-direction, so the states in each layer have a Rashba-type energy term \cite{Rashba}:
\begin{equation}
\delta_{SO}( {\bf{k}}) = \alpha ( \bf{k} \times \hat{\bf{z}} ) \cdot {\bm \sigma} \rightarrow \alpha ( (\bf{k} - \bf{k}_{\mathrm{corner}}) \times \hat{\bf{z}}) \cdot {\bm \sigma}.
\end{equation}
Here, $\bm{\sigma}$ is the spin operator and $\alpha$ gives the strength of this effect; in the original nearly-free electron case \cite{Rashba}, $\bf{k}$ would be proportional to the carrier momentum. For our purposes, a sufficiently good approximation is instead to take $\bf{k}$ relative to the nearest corner of the BZ, since the unreconstructed energy bands form approximately circular energy contours around these points. The sign of the term is opposite in the top and bottom layers. The Rashba term only involves $\sigma_x$ and $\sigma_y$, which can be expressed in terms of eigenstates of $\sigma_z$ using the Pauli spin matrices.  We can write the Hamiltonian matrix including this term by using the layer states labeled by the $z$-component of their spin, giving: 
\begin{widetext}
\begin{equation}
    H_{b+SO}({\bf{k}}) = \begin{bmatrix}
        \epsilon({\bf{k}}) + \mu B_z & \delta_{SO}( {\bf{k}}) & -t_{\perp} & 0 \\
        \delta^{*}_{SO}( {\bf{k}}) & \epsilon({\bf{k}}) - \mu B_z & 0 & -t_{\perp} \\
		-t_{\perp} & 0 & \epsilon({\bf{k}}) + \mu B_z & -\delta_{SO}({\bf{k}}) \\
		0 & -t_{\perp} & -\delta^*_{SO}({\bf{k}}) & \epsilon({\bf{k}}) - \mu B_z
     \end{bmatrix}.
\end{equation}
\end{widetext}
Here, the first two basis states are in the upper half of the bilayer and have up and down spins with Zeeman energy $\pm \mu B_z$ in an applied field $B_z$; these states are coupled by the SO interaction. The other two basis states are in the lower half of the bilayer and have an SO interaction which is opposite in sign to that for the top layer. The (spin-independent) tunnelling terms connect states of the same spin in top and bottom layers, while the spin-orbit terms connect states of opposite spin in the same layer. In zero field, this Hamiltonian gives the $A$ and $B$ state energies, each of which is doubly (Kramers) degenerate due to the overall mirror symmetry of the bilayer:  
\begin{equation}
\epsilon_{A,B}({\bf{k}}) = \epsilon({\bf{k}}) \pm \sqrt{t_{\perp}^2 + |\delta_{SO}({\bf{k}})|^2}.
\end{equation}
Comparison with Eqn.~(3) shows that the spin-orbit term does not mix the $A$ and $B$ bands; it merely increases the $A-B$ energy splitting. In nonzero field, there are four different eigenvalues, which for $\mu B > |\delta_{SO}({\bf{k}})|$ may be labelled approximately as spin up and down in the $A$ and $B$ bands.

Finally, we introduce the CDW perturbation $V$. To first order, it only connects states in the same layer (and of the same spin) that differ by a single CDW $\bf{q}$-vector \cite{SandH2014}. The antisymmetry of this perturbation means that its matrix element between basis states in the upper layer is equal and opposite to that for the lower half of the bilayer, and we may write the values as $\pm V$ respectively. Hence, to describe the hybridisation between two sets of $A$ and $B$ bands, we have the 8 $\times$ 8 Hamiltonian matrix given in Eq.~ (7). (We refrain from reproducing the 16 $\times$ 16 matrix, which is the minimal model to represent all four crossings shown in Fig.~1.) In Eq.~(7), the two 4 $\times$ 4 blocks on the diagonal reproduce Eq.~(5) for the two sets of states connected by a CDW, while the off-diagonal blocks represent the CDW couplings between them. The basal plane wavevector ${\bf{k}}' = {\bf{k}} \pm {\bm{\delta}}_a$ or $\pm {\bm{\delta}}_b$ for hybridisation by an $a$- or $b$-direction CDW. For simplicity, we take the CDW matrix element $V$ to have the same value for the $a$ and $b$ CDW modulations.
 %
\begin{widetext}
\begin{equation}
    \begin{bmatrix}
        \epsilon({\bf{k}}) + \mu B_z & \delta_{SO}({\bf{k}}) & -t_{\perp} & 0 & V & 0 & 0 & 0\\
        \delta^{*}_{SO}({\bf{k}}) & \epsilon({\bf{k}}) - \mu B_z & 0 & -t_{\perp} & 0 & V & 0 & 0\\
		-t_{\perp} & 0 & \epsilon({\bf{k}}) + \mu B_z & -\delta_{SO}({\bf{k}}) & 0 & 0 & -V & 0\\
		0 & -t_{\perp} & -\delta^*_{SO}({\bf{k}}) & \epsilon({\bf{k}}) - \mu B_z & 0 & 0 & 0 & -V\\
		V & 0 & 0 & 0 & \epsilon(\bf{k}') + \mu B_z & \delta_{SO}(\bf{k}') & -t_{\perp} & 0 \\
		0 & V & 0 & 0 & \delta^{*}_{SO}(\bf{k}') & \epsilon(\bf{k}') - \mu B_z & 0 & -t_{\perp}\\
		0 & 0 & -V & 0 & -t_{\perp} & 0 & \epsilon(\bf{k}') + \mu B_z & -\delta_{SO}(\bf{k}') \\
		0 & 0 & 0 & -V & 0 & -t_{\perp} & -\delta^*_{SO}(\bf{k}') & \epsilon(\bf{k}') - \mu B_z
     \end{bmatrix}
\end{equation}
\end{widetext}
\section{$c$-axis coupling}
The quantity $t_{\perp}$ arises from tunnelling between the two halves of a bilayer. It has been stated e.g.~in Refs.~\onlinecite{Andersen, DGA+SC, HandS2012, HRS2015} that $t_{\perp}$ has a strong dependence on $\bf{k}$, and is zero on the nodal line $k_x = k_y$, but this is not the case. If $t_{\perp}$ were indeed zero at the nodal line, Eqn.~(3) indicates that there would be zero $k_z$ dispersion on this line, whereas DFT calculations \cite{Andersen, Efimov, Yelland1} show a nonzero dispersion. In adducing evidence from DFT calculations, we do not rely on their absolute accuracy, merely that if they give finite $k_z$ dispersion, then they show that zero dispersion is \emph{not} imposed by crystal symmetry. The origin of the supposed zero of $t_{\perp}$ is the assumption that the tunnelling between the two halves of the bilayer takes place at the Cu ions, via orbitals of $s$ (or equivalently $d_{3z^2-r^2}$) symmetry \cite{Andersen}. Our own band structure calculations, using the method of Ref.~\onlinecite{Yelland2} show that the tunnelling is dominated by paths via the O-$p$ - Cu-$d_{(x^2 - y^2)}$ conduction band orbitals, because the large O$^{2-}$ ions give a noticeable electron density towards the centre of the bilayer. This main O-Cu tunnelling path is directly in the $c$-direction, between equivalent lobes of wavefunctions on Cu and O, so there is little $\bf{k}$-dependence of $t_{\perp}$ - the \emph{intra}-bilayer tunnelling, and we shall take $t_{\perp}$ as a constant. 

However, the main tunnelling path for the inter-bilayer tunnelling $t_c$ is via a $p_z$-orbital on the apical oxygen. In a tetragonal cuprate, this would lead, for identical reasons, to the expression that has been claimed for $t_{\perp}$:
\begin{equation}
t_c = - \frac{t_{c0}}{4} (\cos k_xa - \cos k_ya)^2.
\end{equation}
This would lead to $t_c = 0$ and again to zero dispersion with $k_z$ on the nodal line, for a tetragonal superconductor, but this result is not exact in orthorhombic YBCO \cite{LiangWheatley} and is not seen in DFT calculations. Nevertheless, the multiple hops for this path, and the effects leading to Eqn.~(8) suggest that $t_c$ will be much smaller than $t_{\perp}$ near the nodal region, and hence at the crossing points where FS reconstruction can occur. These are the reasons for taking $t_c = 0$ in our modelling.
\section{Effects of the CDW - Fermi surface calculations}
We are now in a position to illustrate, by explicit numerical calculations, how various physical effects contribute to the FS reconstruction by the CDWs. In our calculations, we take a double-$\bf{q}$ CDW, with incommensurate $\delta$-values, which for simplicity we set to 0.32 for both components, and take  $\delta$ as independent of field~\cite{Chang2012, Gerber2015, Chang2015}. 
 We take the value $0.2 \times t$ for $t_{\perp}$. This value is chosen for clarity to give well-separated $A$ and $B$ bands; smaller values may be needed to fit the experimental QO frequency splittings, and to be consistent with ARPES observations \cite{Fournier2010} that do not resolve the $A-B$ separation along the nodal direction. For the perturbation due to the CDW we take $V = 0.15 \times t$, giving it a value comparable with the quantity determining the $A-B$ splitting. For the SO interaction, we take a value for $\alpha$ in Eq.~(4) of $0.1 a \times t$. This was chosen to be sufficiently large to make the SO hybridisation gaps easily visible. We have expressed all energy variables in terms of $t$, so that our calculations of FS areas are independent of the value of $t$, which then determines the effective masses associated with these areas. 

The combination of the CDW $\delta$-values and the positions of the unperturbed Fermi surfaces determine the areas of the reconstructed electron pockets. The value of $\epsilon_F$ was chosen to give pocket areas which are close to the 2\% of the BZ given by QO measurements. This implies Fermi arcs which are close to the antiferromagnetic BZ boundaries. This is consistent with ARPES observations in underdoped YBCO \cite{Fournier2010}. Alternatively, the effects of underdoping or PG may reduce the values of $t'$ and $t''$ below those of Ref.~\onlinecite{Andersen}. A reduction by a factor of 2 would give positions of the Fermi arcs in agreement with experiment. A simplistic picture ignoring the PG and constructing an unperturbed FS shape containing $\sim$ 1.1 holes (doping $p \sim$ 0.1) and $\delta \sim$ 0.32 would give FS pockets considerably smaller than those observed.

Typical results of numerical diagonalization of the full 16 $\times$16 Hamiltonian matrix, calculated at zero $B_z$ are shown in Fig. 2. In (a, c) we give the FS shapes in the absence of the CDW, to make it clear in (b, d) which states are mixed by the CDWs. In (b) we see a smaller FS sheet that is $A$ in character plus a larger one that is mainly $B$ in character; however (d) shows that the $A - A$ splitting, which gives the smaller FS in (b), only arises because of the weak SO interaction. Without it, we would obtain intersecting Fermi surfaces like those depicted in Fig. 1 (e), (f).
\begin{figure}[t]
\begin{center}
\includegraphics[width=0.95\linewidth]{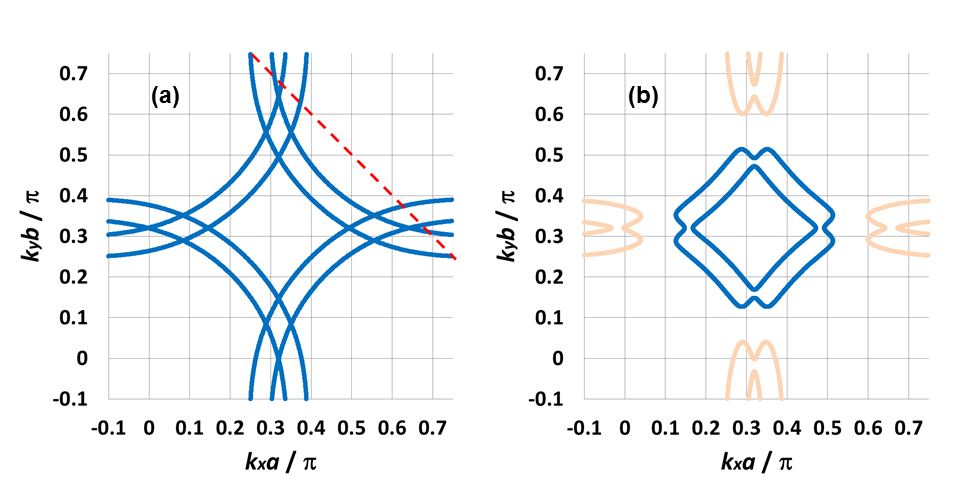}
\includegraphics[width=0.95\linewidth]{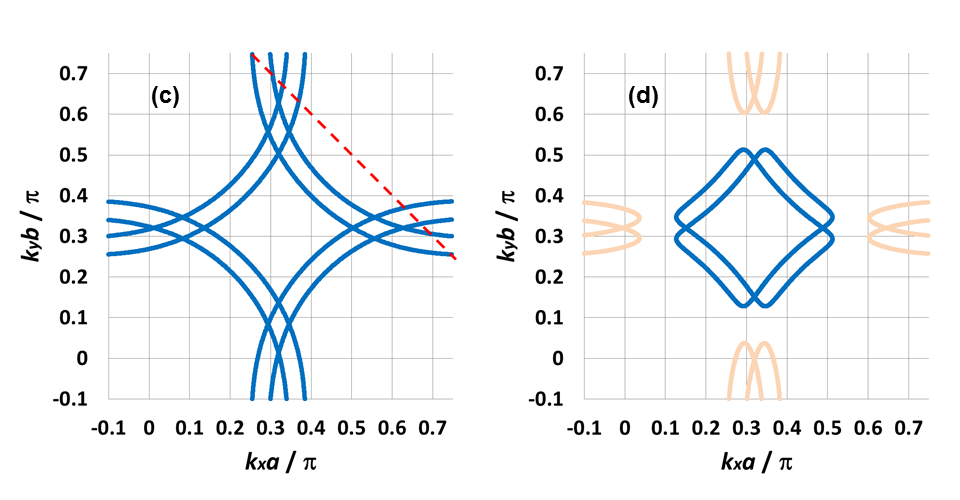}
\end{center}
\caption{Reconstructed Fermi surfaces in zero magnetic field created by the CDWs. In (a) are shown the bands in the absence of hybridisation by the CDWs. The antiferromagnetic zone boundary is marked by the diagonal red dashed line. In (b) are shown the Fermi surfaces as reconstructed from the Fermi arcs by the CDWs. We assume that the paler parts are removed by the PG. (c) and (d) are calculated with the SO interaction set to zero. By comparison of (b) with (d) one can see which crossings are turned into avoided crossings by the SO interaction. Careful comparison of (a) with (c) shows that the SO interaction gives a greater $A - B$ separation, as expected from Eqn. 6.}
\end{figure}
\section{Effects of magnetic field and SO interactions on Fermi surface reconstruction}
The Zeeman energy of the electron spins at the fields used for QO measurements is substantial. It is comparable to the Landau level spacing and there are only $\sim$ 10 Landau levels below the Fermi level in the small electron pockets. In Fig.~3 we show that in addition, the effects of magnetic field along the $c$-direction are non-trivial. Fig 3 (a) shows the effects of the Zeeman energy (set to a value $\pm 0.1 \times t$) on the bands in the absence of hybridisation, with an expanded view at a single crossing in (b). The effects of the CDWs are shown in (c) (d). The spin-split Fermi surfaces are labelled with arrows representing the spin directions. Two interesting effects are revealed in (c) (d). It will be noticed that there are places where Fermi surfaces cross each other without any splitting. At the high magnetic fields employed for QO measurements, it is likely that any residual gaps in these regions due to higher order effects will be ineffective and the electron orbits will tend to pass pass  through them. Another very important qualitative feature of the SO interaction is revealed by tracing the spin directions from regions well away from the crossing region. 
\begin{figure}[t]
\begin{center}
\includegraphics[width=0.95\linewidth]{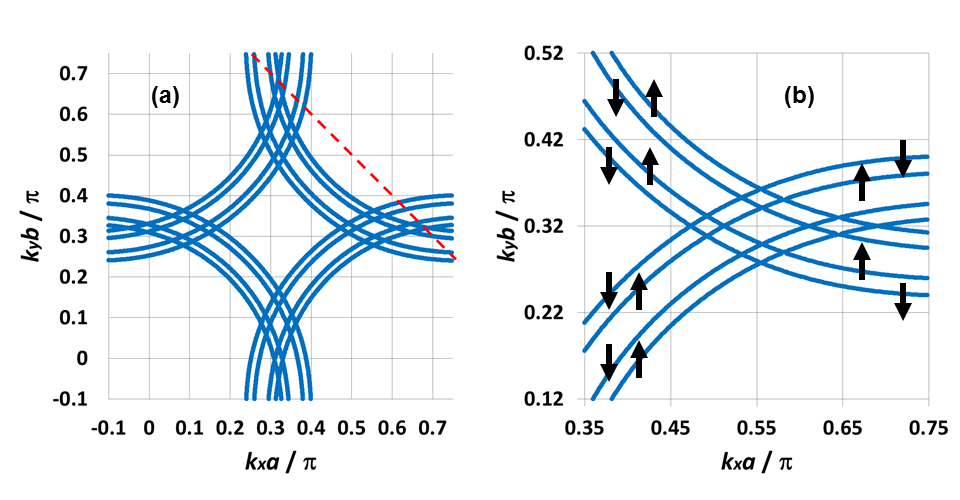}
\includegraphics[width=0.95\linewidth]{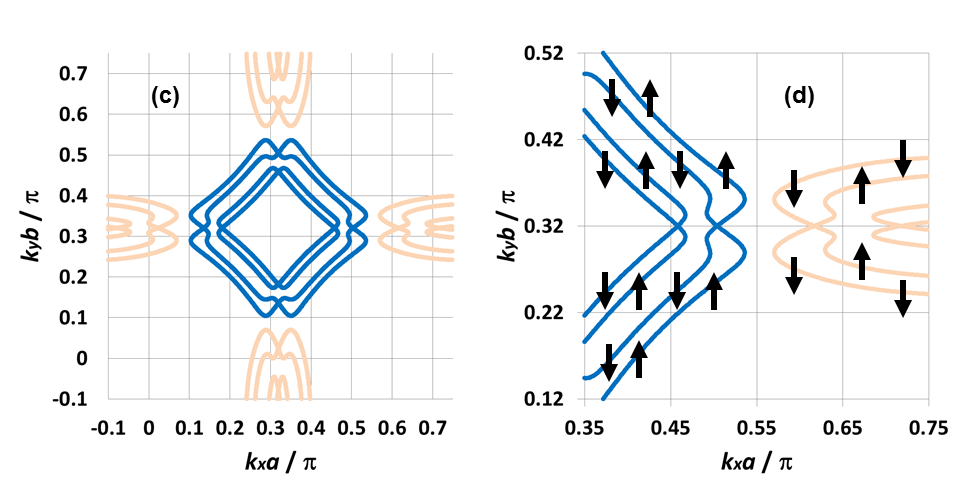}
\end{center}
\caption{Reconstructed Fermi surfaces in a magnetic field and details of crossings. (a) shows the spin-split bands in the absence of hybridisation by the CDWs, and (b) gives a close-up of one of the hybridisation regions for comparison with (c) (d). The spin directions in the various bands are marked by arrows. (c) and (d) show the same regions with the avoided crossings and spin flips caused by the CDW.}
\end{figure}
We find in those places where the SO interaction enables an avoided crossing, that the carrier spin direction is flipped as well. As emphasised in Ref.~\onlinecite{HRS2015}, a carrier going around an orbit with spin flip can show an anomalously small average Zeeman energy or $g$-factor. In all the orbits shown in Fig 3 (c), there is an even number of spin flips, so the carriers should have close to zero average Zeeman energy.  However, this conclusion is modified if the electron orbits can pass by `magnetic breakdown' through the SO gap or other small gaps. This process can alter both the areas of the orbits and the spin energy, so we now consider it.
\section{Magnetic breakdown}
The representation of the motion of a carrier in a lattice in a magnetic field by the progress of its $\bf{k}$-vector around a Fermi surface is an approximation. If the carrier comes to a small `forbidden' region created by an energy gap, it is possible for the particle to `tunnel' through the forbidden region, particularly at high magnetic fields. This is called magnetic breakdown. If two FS sheets approach each other but have a gap between them in $\bf{k}$-space of value $k_g$, then the probability $P$ of tunneling across the gap is given by \cite{Fortin}: 
\begin{equation}
P \approx \mathrm{e}^{-k^2_g \ell^2_m}
\end{equation}
where $\ell_m$ is the magnetic length in a field $B_z$, defined by $\ell_m^2=\hbar/eB_z$. For the case considered here, the important process is magnetic breakdown from $A$ to $A$ or $B$ to $B$ at the red crossing points in Fig. 1. (We propose that breakdown out of the Fermi arcs towards the edges of the BZ is prevented by the PG.) Assuming that the CDW gaps are comparable with those due to $A-B$ splitting, we may use the QO data to give an order of magnitude for the gap $k_g$ between different FS areas, and this will indicate (independent of our calculations above) at what fields magnetic breakdown may be important. The QO frequencies in YBCO are around 2\% of the BZ area, i.e.~$\sim 6 \times 10^{18} m^{-2}$ and the spread of frequencies is $\sim 40\%$, giving a gap $k_g$ between the corners of two concentric areas $\sim 2.5 \times 10^8 m^{-1}$. Now we may write the probability of magnetic breakdown in the form:
\begin{equation}
P \approx \mathrm{e}^{-B_0 / B_z}, \qquad \mathrm{where} \qquad B_0 = \frac{\hbar k_g^2}{e}
\end{equation}
Using our rough estimate, of $k_g$, we obtain $B_0 \sim$ 40 T, which is in the range where QO measurements are carried out, and suggests that magnetic breakdown may well be dominant at the highest fields. Indeed, according to our calculations, the hybridisation gaps between bands of like character have small SO values at the crossing points. Hence, magnetic breakdown will play an important role even at lower fields, and it will give a high probability that the ${\bf{k}}$-vectors of the carriers pass straight through these crossings. 
\section{Effects of tilting the magnetic field away from the c-axis}
An important experimental variable in QO measurements is to change the direction of the applied field. In two-dimensional materials, this allows the comparison of the Zeeman and Landau level spacings, which have a different dependence on angle and can lead to `spin zeroes' in the QO signal for particular angles between {\bf{B}} and the {\bf{c}}-axis \cite{Ramshaw2011,SandH2014}. There is a further effect of tilting the magnetic field: $B_0$ in the expression for magnetic breakdown becomes $B_0 / \mathrm{cos}(\theta)$, when the field is tilted by an angle $\theta$ from {\bf{c}}, so the probability of magnetic breakdown is decreased. In the present case, there is an additional interest: the SO-term in Eqn. 4 behaves like a magnetic field in the basal plane, parallel to the unreconstructed Fermi surface and of opposite sign in the two halves of a bilayer. This can interfere with a real magnetic field, which when tilted gives a basal plane component equal in the two halves of a bilayer.

 In Fig. 4 we show the effect of \emph{field angle} only; we have neglected the weak variation with angle of the electronic $g$-factor (also a SO effect), because it does not give rise to qualitatively new behaviour. It will be seen in Fig. 4 (a) (b) that a field tilted towards the $y$-direction removes the SO spin flips for the crossings in the $k_x$ direction and replaces them with small gaps, but the spin-flips remain for the $k_y$ crossings. This is confirmed by Fig. 4 (d), which shows the effect of tilting the field direction towards {\bf{x}}. When the field is tilted along the $xy$-direction, both SO spin flips are removed and replaced with somewhat smaller gaps, as shown in Fig. 4 (e) (f). These gaps arise when the sum of the basal plane applied field and SO effective field further breaks the bilayer symmetry, so that more CDW hybridisation matrix elements become nonzero.

Thus, tilting the magnetic field will introduce a fourfold anisotropy in the values of the gaps and hence in the connectivity of the electron orbits. Our model can therefore provide a framework for understanding the fourfold variation seen in QO measurements in tilted fields \cite{SandH2014}. This depends on the topology of the orbits given by magnetic breakdown, and we now consider this matter in detail.
\begin{figure}[t]
\begin{center}
\includegraphics[width=0.95\linewidth]{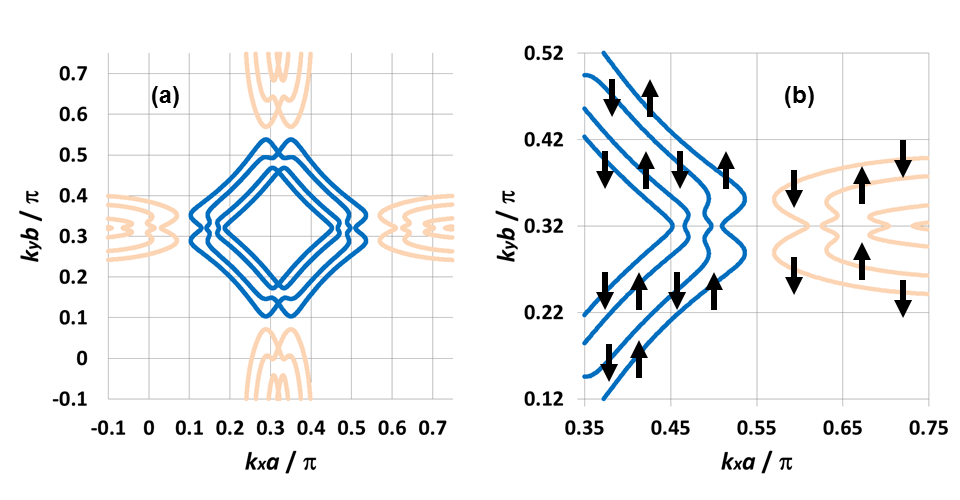}
\includegraphics[width=0.95\linewidth]{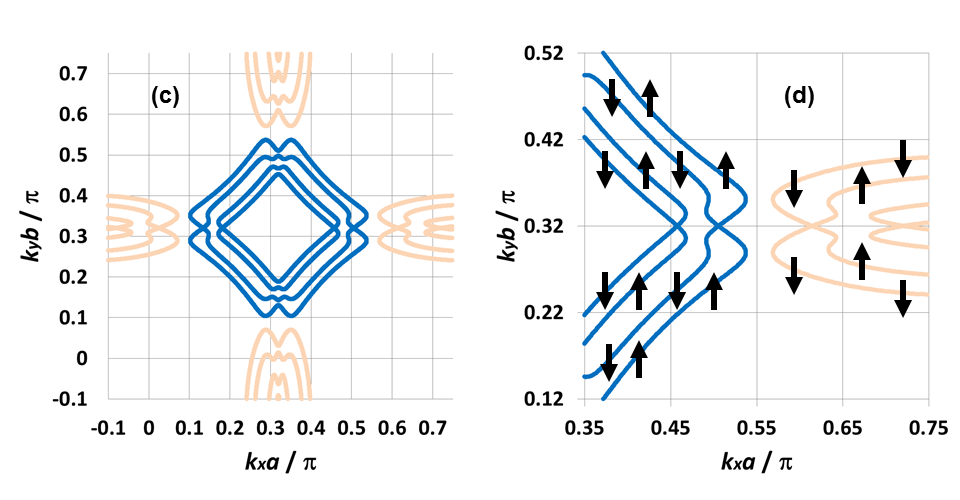}
\includegraphics[width=0.95\linewidth]{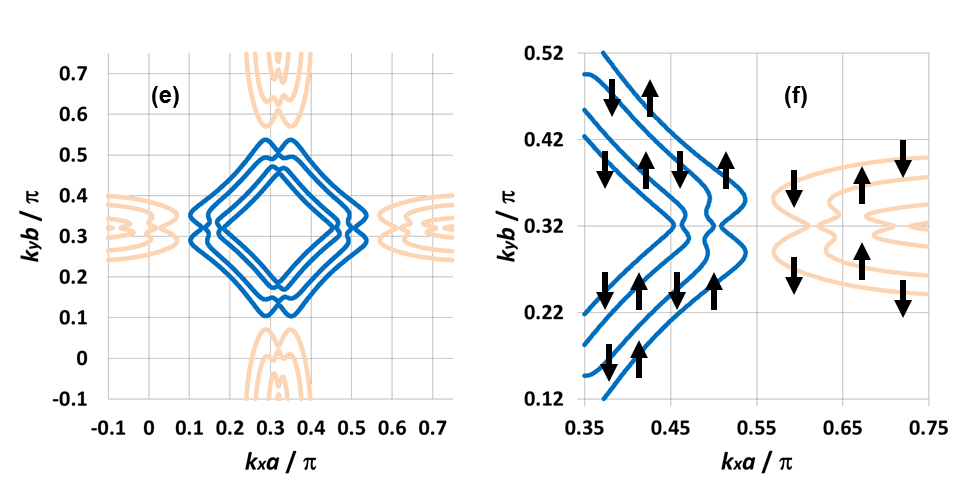}
\end{center}
\caption{Effects of tilted magnetic field and SO interaction on Fermi surface reconstruction. (a) The same magnitude of field as in Fig. 3 is applied at an angle of 45$^{\circ}$ to {\bf{c}}, tilted along the $y$-direction. (b) shows in detail the removal of spin-flips and their replacement by small gaps for the crossings near the $k_x$-direction. (c) and (d) are for the field tilted towards {\bf{x}}, and confirm that the SO-induced spin flips are not removed for the crossings close to the field-tilt direction. If the field is tilted towards the $xy$-direction, smaller gaps are created at both crossing points, as indicated in (e) (f). }
\end{figure}
\section{Magnetic breakdown and orbit areas}
\begin{figure}[t]
\begin{center}
\includegraphics[width=0.95\linewidth]{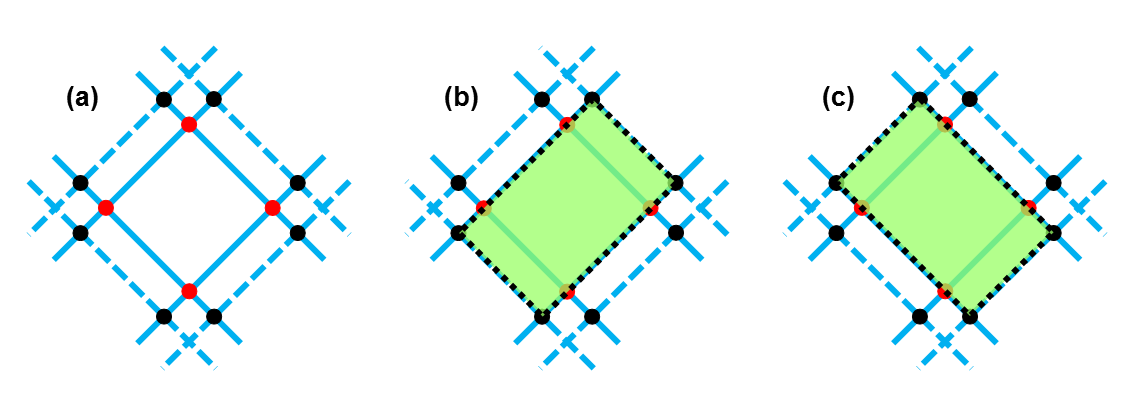}
\includegraphics[width=0.95\linewidth]{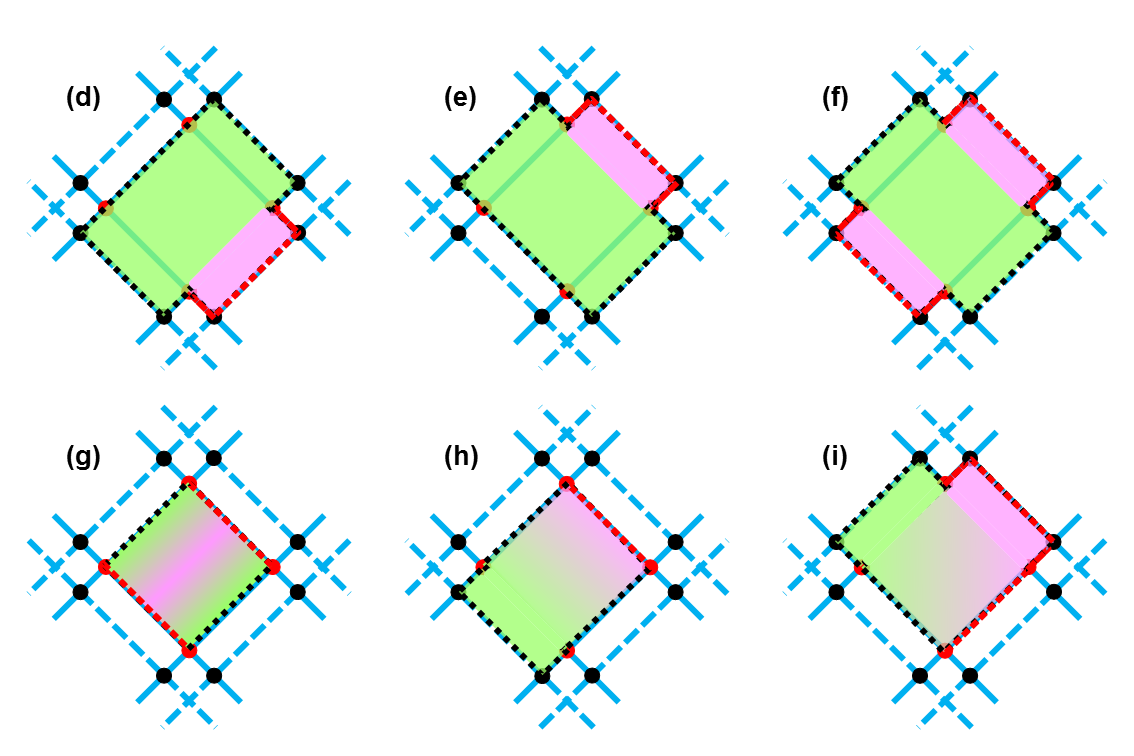}
\end{center}
\caption{Fermi surface orbits, magnetic breakdown and spin-flip
The straight lines in {\bf{k}}-space represent schematically the $A$ and $B$ (dashed) bands that intersect to form the small electron pocket Fermi surfaces. Strong hybridisation (which would round the corners) and reflection is expected at the black points marking $A - B$ crossings. Weak hybridisation, spin-flip on reflection and a larger probability of transmission is expected at the red $A - A$ crossing points. The shaded areas represent the different kinds of possible {\bf{k}}-space closed orbits of carriers in a magnetic field which would be detected by QO measurements. The change in colour of the orbit from black to red and back represents a flip of the spin of an electron by the SO interaction.}
\end{figure}
In Fig. 5, we give a schematic representation of the various {\bf{k}}-space orbits predicted by this model. The crossings of the $A$ and $B$ (dashed) bands are shown in (a). The black spots on the $A - B$ crossings represent the strong hybridisation by the CDWs which does not lead to electron spin flip. The red spots mark weak hybridisation by the CDW in the presence of the SO interaction. If a carrier is reflected by the small gaps at these points, it may undergo a spin flip. If it is transmitted, no spin flip occurs. In (b) (c) the shaded areas are the two orbits of medium area, which would represent the central peak in a typical QO spectrum. This peak is dominant in the experimental data, and is also known to have an electron $g$-factor close to the standard value \cite{Ramshaw2011,SandH2014}. In our model, these orbits have no spin flips, which agrees with the observed $g$-factor value. These orbits arise only because of magnetic breakdown, so we deduce that the gaps at the red points are small. 

In (d) - (i), we represent the other possible orbits allowed by the model. In (d) (e) are represented two different orbits of identical area but different orientations; in later panels, we give only the topology, not all possible orientations. The spin flips are represented by using a black dashed line for paths with one spin orientation and a red one for the other. It will be noted that in  orbits (d) to (i) there is some cancellation of the electron spin direction. This is a phenomenon pointed out by Ref. \onlinecite{HRS2015} and gives a reduced effective $g$-factor, and less dependence of orbit area on the value of magnetic field. It is found experimentally that some of the QO sidebands have different spin zero positions from the main frequency. This is an expected consequence of our model. There are in fact three orbits, (f), (g) and (i) which have a $g$-factor close to zero and (i) has an area similar to (b) and (c). When magnetic breakdown is strong, we expect that orbits with many reflections, (f) and (g), will have low weight.

These are the predictions of the model for {\bf{B}} parallel to {\bf{c}}. As shown earlier, tilting the field  away from {\bf{c}} will not only increase the probability of reflection at some of the red crossing points, it can also replace spin-flips with energy gaps at various places, depending on the azimuthal angle of the tilt relative to the {\bf{a}}, {\bf{b}} crystal axes. In particular, we find that orbits (b) and (c) of Fig. 5 acquire additional gaps, which can alter the probability of transmission or change the spin energy on reflection. This, and similar effects on the orbits of type (i), which have the same area, could account for the reported fourfold azimuthal dependendence of the QO signal which may arise from these orbits \cite{SandH2014}. Our model clearly predicts a wealth of different effects and gives a scenario capable of accounting for the multiple pocket areas given by QO measurements, although a detailed fit of the extensive existing data \cite{QO2007,SandH2014, Riggs2011,SHL2012,DL2015,QO-1,QO-2,QO-3,QO-4,QO-5,QO-6,QO-7,QO-8,QO-9,QO-10,QO-11} is beyond the scope of the present Paper.
%
%
\section{Discussion}
The model we have described gives a coherent account of the influence of the CDW on the QO spectra observed in underdoped YBCO.  We have ignored all effects of the CuO chains, since they appear to be non-metallic in the underdoped region, and the different orderings of the occupied chains at different dopings appears to have no effect on the smooth variation of QO frequency with doping \cite{QO-10}. We have relied on the pseudogap and underdoping to remove chain and antinodal carrier states which do not appear in QO, Hall, and heat capacity measurements at high field \cite{Riggs2011}, and leave the `Fermi arc' states which are hybridised by the CDWs.  

Recent QO measurements \cite{DL2015} have suggested the existence of an additional hole-like area, approximately $1/5$ of those discussed here, and it was proposed \cite{Psgap} that the pseudogap can also explain this result, which does not come out of our calculations. However, it seems likely that these experimental results arise either from mixing of QO frequencies, which can arise in two-dimensional materials from chemical potential oscillations \cite{QO-8} or `Stark interference' \cite{Stark} corresponding to effects on transport properties of the difference between two areas traversed by electrons \cite{HRS2015}.


A notable feature of the CDW state in YBCO - and in other materials - is that the $\bf{q}$-vector of the CDW appears not to be a nesting vector between two parallel sheets of Fermi surface. It is clear that the nesting argument, which is persuasive in 1-D does not necessarily apply in 2- and 3-D \cite{Johannes & Mazin, Zhu}. The question then remains: what is the driving force for the CDW, and what determines its $\bf{q}$-vectors?

The picture we have of FS reconstruction in YBCO should also be tested against the field-dependence of the Hall effect \cite{LeBoeufHall2007, GrissarXiv}. It has been pointed out \cite{HSarXiv} that the positive and negative curvatures of the reconstructed pockets give rise to a field-dependence of the Hall effect, with a large value and electron-like sign at high fields. The \emph{magnitude} of the high-field Hall coefficient is simply related to the number of carriers, and two pockets each occupying  $\sim$ 2\% of the BZ would give a value $\sim -14$ mm$^3$/C. Earlier data \cite{LeBoeufHall2007} gave a value much larger than this, corresponding more closely to a single electron pocket rather than the two we obtain from calculation. However, more recent data from the same group \cite{GrissarXiv} is in good agreement with expectations. 

We have used ARPES measurements to indicate those portions of the Fermi surface which remain after the effects of the pseudogap and therefore can be reconstructed by the CDW.  However, although CDWs exist in zero magnetic field, there have been no reports of a reconstructed FS observed by ARPES. QO measurements are a bulk effect, and it is possible that the immediate surface - to which ARPES is sensitive - does not support a strong enough CDW. It is also possible that the pocket signal is too weak in ARPES or would only appear at high fields where QO measurements are performed and ARPES measurements are impossible. 
\section{Conclusions}
In conclusion we have demonstrated how multiple QO frequencies observed in YBCO are naturally explained by the odd symmetry of the CDW order, combined with the effects of spin-orbit interaction and magnetic breakdown. Our minimal model for an isolated bilayer gives an account of the main features expected for the reconstructed part of the Fermi surface in underdoped YBCO and leads to a rich phenomenology. Our calculations have been carried out for reasonable values of the parameters to illustrate the effects that are produced in the published quantum oscillation data, including the fourfold anisotropy present when the field is tilted away from the $c$-axis. It is interesting that the tilted field not only investigates the electronic structure which was present with field perpendicular to the bilayers, but also modifies it. It remains for detailed fits, which will need to include the effects of magnetic breakdown, giving a field-dependence to the amplitudes of the various QO frequencies, to establish whether the role of the CDW in Fermi surface reconstruction in YBCO can be fully established. 
\begin{acknowledgments}
We acknowledge the UK EPSRC for funding under grant numbers EP/J016977/1 and EP/J015423/1 and helpful discussions with Mike Gunn,  Suchitra Sebastian and Tim Ziman.
\end{acknowledgments}

\end{document}